\documentclass[preprint,showpacs,preprintnumbers,amssymb]{revtex4}

\usepackage{graphicx}
\usepackage{dcolumn}
\usepackage{bm}
\usepackage{cd}
\usepackage{amsfonts}
\usepackage{amsmath}
\usepackage{times}
\usepackage[T1]{fontenc}
\usepackage[applemac]{inputenc}
\usepackage{fancyhdr}

\renewcommand{\d}{{\rm{d}}}

\begin{document}

\preprint{APS/123-QED}

\title{Survey of roughness by stochastic oscillations}

\author{\textbf{E. Moreau }\\
    \emph{\small{Laboratoire des Sciences des Procédés Céramiques et 
    Traitements de Surface\\ ENSIL, 16 rue d'Atlantis 87068 Limoges, France}}}

\begin{abstract}
In this paper, connection between surface roughness and directed polymers in 
random medium are studied, when the surface is considered as a directed 
line undergoing stochastic oscillations. This is performed by studying the 
influence of a stochastic elastic forcing term $-\kappa y+\eta(s)$, on 
a particle moving along a rough surface. Two models are  proposed and analysed 
in this way: the random-walk process (RW) in its discrete and continuous form, 
and a Markovian process via the Ornstein-Uhlenbeck (O-U) process. It is shown 
that the continuous RW leads to an oscillator equation, via an effective action 
obeying a KPZ equation which is solved analytically. The O-U process allows 
to obtain information on the profile of surface for a long size substrate. The 
analogy with the roughness is achieved by introducing a quantity 
suited to directed line formalism: the height velocity variation 
$\partial h/\partial s$. 
\end{abstract}


\maketitle

\date{\today}
\section{Introduction}

Characterization of surface roughness or interface growth is of 
first importance in many applications: microelectronics, preparation 
of catalysts, magnetic materials (see {\it{e.g.}} \cite{ex1}-\cite{ex4}). 
And numerous models based on numerical as well as analytical approaches 
have been proposed over past decades to treat interface properties. These 
works concern in particular the KPZ equation (scale laws, roughness, growth) 
\cite{kpz1}-\cite{kpz11}, and models treating connections between surface 
growth and directed polymers in random medium \cite{ham1}-\cite{ham7}. The 
present paper focus on the second class of these models, considering the 
roughness of a surface as the result of specific stochastic processes, and 
based on the directed line formalism. This formalism, much used in polymer 
physics \cite{dp1}, allows to analyse the motion of a system by the way of 
transversal and curvilinear coordinates \cite{dp2}-\cite{dp4} (respectively 
noted $y$ and $s$ in this paper). And we propose models showing the analogies 
between the profile of a rough surface and the study of the transversal position 
equation. Note that, in this formalism, $s$ plays the role of the time. For 
example $\d y/\d s$ designs the transversal velocity. Moreover, this formalism 
is also linked to the statistical physics: the partition function 
of the system and its links with the KPZ or the Burgers 
equation, which follow from this formalism, represent useful tools 
\cite{fp1}-\cite{fp3}. Concerning the roughness analysis, different works 
have treated this problem \cite{fp1}, \cite{r1}-\cite{r5}, but the roughness is 
nevertheless poor studied from a purely stochastic point of view. Among recent 
works treating stochastic surface properties, we can cite \cite{markov1} 
and \cite{markov2} which both have a Markovian approach, using a Langevin 
equation with a stochastic noise.

\begin{figure}[!h]
	 \centerline{\includegraphics[scale=0.6]{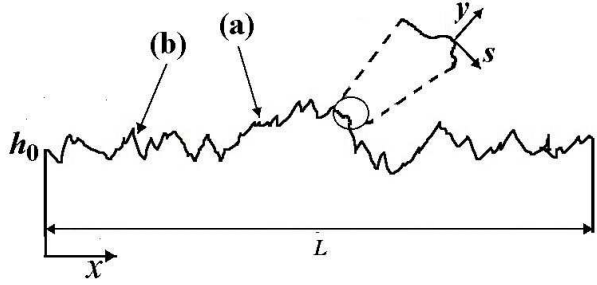}}
	 \caption{{\small{Profile of surface: result of the 
	 conflict between the random medium and a directing strength 
	 $-\partial\Phi/\partial y$. The local slope (a) (resp. (b)) stands 
	 for a weak (resp. high) value of the hvv. }}}
	 \label{rw2}
\end{figure}

In the models presented here, we assumed that the profile of 
surface is described by the trajectory of a particle undergoing a 
directing stochastic strength in addition to diffusion. We will use for 
this purpose the elastic forcing term 
$f(y,s)=-\kappa y+\eta(s)$ inducing stochastic oscillations, or 
equivalently, a potential $\Phi$, such as $f=-\partial\Phi/\partial 
y$; $\eta$ having for example the properties of a Langevin force. 
For deposition processes, we suppose that the 
surface start at the height $h_{0}$ , and we are interested in the height 
velocity variation (hvv), $\partial h/\partial s$, which gives local 
information on the structure of the surface to analyse: for a fast 
variation of the height, we will have a high value of the hhv, 
(Fig. (\ref{rw2})). Hence, this quantity is the reflect of local 
irregularities. But, the hvv do not have to be confused with the quantity 
$\partial h/\partial x$, where $x$ stands for the horizontal coordinate: 
locally, the slope $\partial h/\partial x$ may takes an infinite value 
({\it{i.e}} parallel to the $h$ axis), whereas 
$\left(h(s_{1})-h(s_{2}))/(s_{1}-s_{2})\right)$ is weak.

The outline of the paper will be as follow: we first derive properties related 
to profile of surface from pure and forced Random Walk (RW). In what 
follows, we will employed the term forced RW to design a RW undergoing a 
forcing term in addition to the diffusion. Then, we justify the use of an 
oscillator model comparing the directed line to a string vibrating. This will 
allow to use the restricted partition function to determine the transversal equation of motion, in presence of a quadratic 
stochastic potential. In the last part, we shall concentrate on the O-U 
equation of motion for the transversal position, and particularly on the 
asymptotic solutions for $s\to\infty$.

\section{Topology of surface view as a Random Walk process}

The models presented here concern the use of the directed line formalism 
associated to a RW process. In its discrete form, we based our approach on 
the paper of Vilgis \cite{dp1}, treating directed polymers in random media. 
Our derivation is somewhat different of Vilgis, but leads to the 
well-known partition function of polymer physics as showed later on. 
\subsection{Discrete random walk process}

\begin{figure}[!h]
	 \centerline{\includegraphics[scale=0.5]{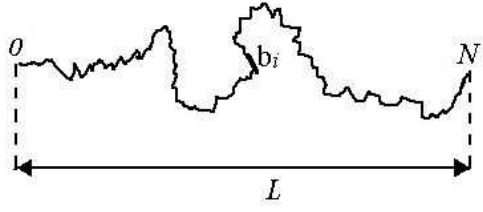}}
	 \caption{{\small{Random Walk: $N$ steps of length $b$ are required 
	 in order to cover the surface on a linear size substrate $L$.}}}
	 \label{graph1}
\end{figure}

We start from a set of vectors $\{\mathbf{b}_{i}\equiv
b_{i}\,\mathbf{e}_{s};\,i=1,\ldots , N\}$, which are successively 
connected, and recovering the whole of the surface (Fig. 
(\ref{graph1})). We suppose they are statistically independent.

The probability to find such a set is given by:
\begin{equation}
\label{proba}
P(\{\mathbf{b}_{i};i=1,\ldots N\})=\prod_{i=1}^{N}
p(\mathbf{b}_{i}),
\end{equation}
where the probability $p(\mathbf{b}_{i})$ corresponds to a step of
the random walk of $\mathbf{b}_{i}$,
with the mean value properties:
\begin{equation}
     \langle \mathbf{b}_{i}\rangle=0\qquad\langle
\mathbf{b}_{i}\mathbf{b}_{j}\rangle=b^{2}\delta_{ij}.
\end{equation}
$\delta_{ij}$ is the Kronecker symbol and $b$ is a characteristic
length of the shift of position of the surface from one to the other.
For the step $\mathbf{b}_{i}$, a normalized gaussian probability law is 
taken:
\begin{equation}
p(\mathbf{b}_{i})=(1/2\pi b^{2})^{1/2}
\exp(-\mathbf{b}_{i}^{2}/2b^{2}),
\end{equation}
and the associated distribution of the surface length reads
\begin{equation}
     P(\mathbf{R})=\int \d\mathbf{b}_{1}\ldots\int\d\mathbf{b}_{N}\;
\delta\left( \mathbf{R}-\sum_{i=1}^N \mathbf{b}_{i}\right)\prod_{i=1}^{N}
p(\mathbf{b}_{i});
\end{equation}
the vector $\mathbf{R}=\sum_{i=1}^N \mathbf{b}_{i}$ characterizing the line. 
Then, using the Fourier transform representation of the Dirac delta, a Gaussian 
law for the length distribution arises quite naturally
\begin{equation}
     P(\mathbf{R})=(1/2\pi b^{2}N)^{1/2} \; \exp(-
\mathbf{R}^{2}/2 b^{2}N).
\end{equation}
And from this expression, the quadratic mean ``length'' of the surface  
can be evaluated:
\begin{equation}
\langle \mathbf{R}^{2}\rangle=\int
\mathbf{R}^{2}\;P(\mathbf{R}^{2})\,\d\, \mathbf{R}= b^{2}N.
\end{equation}
Note that the quantity $\langle \mathbf{R}^{2}\rangle$ preserves 
information on the length of the trajectory, contrary to the 
geometrical quantity $\mathbf{R}$. However, this result is a limiting case of 
a pure RW trajectory since it implies that the entanglements, and so the 
length of the surface, could escape to infinity. Thereby, for this case we  
label $N$ such as 
\begin{equation}
    \label{a3}
    \langle{\bf{R}}^{2}\rangle_{{\rm{rw}}}=b^{2}N_{{\rm{rw}}},
\end{equation}
with the meaning that $N_{{\rm{rw}}}$ steps are necessary in average 
for a pure RW to describe the profile of surface. The other limiting case, 
\textit{i.e.} the straight line (sl), is given by
\begin{equation}
    \label{a4}
    \langle{\bf{R}}^{2}\rangle_{{\rm{sl}}}=b^{2}N^{2}_{{\rm{sl}}}\simeq L^{2},
\end{equation}
in such a way that the general case (g), representing a forced RW, and
included between the cases (\ref{a3}) and (\ref{a4}) , may be written
\begin{equation}
    \label{a5}
    \langle{\bf{R}}^{2}\rangle_{{\rm{g}}}=b^{2}N^{2\gamma}_{{\rm{g}}};
    \qquad N_{{\rm{rw}}}\le N_{{\rm{g}}}\le N_{{\rm{sl}}},
\end{equation}
where $\gamma$ is a critical exponent representative of a strength 
acting on the RW. In addition, we have the 
conditions
\begin{equation}
    \label{cond1}
    b^{2}N^{2\gamma}_{{\rm{g}}}\to b^{2}N^{2}_{{\rm{sl}}}\qquad
    {\rm{for}}\qquad N_{{\rm{g}}}\to N_{{\rm{sl}}};
\end{equation}
which implies $\gamma=1$, and  
\begin{equation}
    \label{cond2}
    b^{2}N^{2\gamma}_{{\rm{g}}}\to b^{2}N_{{\rm{rw}}}\qquad 
    {\rm{for}}\qquad N_{{\rm{g}}}\to N_{{\rm{rw}}}; 
\end{equation}
implying $\gamma=1/2$. Thus, $1/2<\gamma<1$, available for a line moving in 
the plane $(h,x)$. This result may also be expressed in term of the length 
$l\simeq\sqrt{\langle{\bf{R}}^{2}\rangle_{{\rm{g}}}}$ of the surface:
\begin{equation}
    \label{a6}
    l=b\;{N_{{\rm{g}}}}^{\gamma}.
\end{equation}
In the case of a two-dimensional RW, $\gamma$ may takes the form 
$\gamma=1/d_{{\rm{f}}}$, where $d_{\rm{f}}$ designs the fractal 
dimension of the directed line. As a matter of fact, $d_{\rm{f}}$ describes 
the degree of irregularity of the surface, and for $\gamma= 1$ 
and $\gamma= 1/2$, $d_{\rm{f}}$ becomes identical to the topological 
dimension of the system. Clearly, the exponent $\gamma$ is linked to the 
number of steps $N_{{\rm{g}}}$. A relation between the two quantities may be 
obtained from the probability of the general case, $P_{{\rm{g}}}({\bf{R}})$, 
when the line undergoes both effects of diffusion and of an external force. 
Anticipating the continuum approach described afterwards, we write this 
probability
\begin{equation}
    \label{a7}
    P_{{\rm{g}}}({\bf{R}})=\mathcal{N}
    \exp\left(-{\bf{R}}^{2}/2b^{2}N_{{\rm{g}}}-\beta\Phi({\bf{R}})\right),
\end{equation}
where $\Phi({\bf{R}})$ stands for an external potential. For example, 
in the case of a harmonic potential, $\Phi=(\kappa/2){\bf{R}}^{2}$, the 
quadratic mean length is
\begin{equation}
    \label{a8}
    \langle \mathbf{R}^{2}\rangle_{{\rm{g}}}=\int
    \mathbf{R}^{2}\;P_{{\rm{g}}}(\mathbf{R}^{2})\,\d\, \mathbf{R}=
    \frac{b^{2}N_{{\rm{g}}}}{1+b^{2}\beta\kappa N_{{\rm{g}}}}
\end{equation}
which leads, with Eq. (\ref{a5}), to the expression
\begin{equation}
    \label{a9}
    \gamma=\frac{\ln\left(N_{{\rm{g}}}/(1+b^{2}\beta\kappa N_{{\rm{g}}})\right)}
    {2\ln\left(N_{{\rm{g}}}\right)}.
\end{equation}
One can then obtain information on the topology of a surface, in the sense 
that the ratio $l/L$ is the reflection of its irregularities and roughness. 
The fractal dimension may be evaluated experimentally, for example, by the 
box-counting method with boxes of length $b$ (in the same way as for 
electric discharges \cite{frac}). In simulations, the varying parameter 
would be the number of steps of the RW.
\subsection{Equations of evolution}
If we express now the step as
$\mathbf{b}_{i}=\mathbf{R}_{i}-\mathbf{R}_{i-1}$, and writing shortly
$P=P(\{\mathbf{b}_{i};i=1,\ldots N\})$,
the probability (Eq. (\ref{proba})) may be  rewritten  
\begin{equation}
P=\left(\frac{1}{2\pi b^{2}}\right)^{N/2}
\exp \left[-\frac{1}{2b^{2}}\sum_{i=1}^{N}
(\mathbf{R}_{i}-\mathbf{R}_{i-1})^{2}\right].
\end{equation}
One can then associated to the exponential argument a symbolic 
Hamiltonian 
\begin{equation}
    \label{oub1}
    \beta H_{0}=\frac{1}{2b^{2}}\sum_{i=1}^{N}
    (\mathbf{R}_{i}-\mathbf{R}_{i-1})^{2}.
\end{equation}
This is the energy of a mass-spring chain of a one dimensional lattice.
This suggest to take the continuous limit of this discrete chain:
$\bf{R}_{i}-\bf{R}_{i-1}\to\partial{\bf{R}}/\partial s$, 
and $\sum_{i=1}^{N}\to\int_{0}^{N}$, where $s$ is the 
dimensionless curvilinear coordinate. In this limit, the probability of 
the pure RW reads
\begin{equation}
    P\;\to\;\mathcal{N}
    \exp \left[-\frac{1}{2b^{2}}\int_{0}^{N}
    \left(\frac{\partial{\mathbf{R}}}{\partial s}\right)^{2}\,\d s\right],
\end{equation}
where $\mathcal{N}$ is a normalization constant.
 From this result, one can construct the partition function as a sum
over all the possible paths starting from $s=0$ and reaching the 
linear size substrate at $s=N$:
\begin{equation}
    Z=\mathcal{N}\!\!\!\!\!\!\!\sum_{\mathrm{All\;paths}\;\mathbf{R}(s)}
    \!\!\!\!\!\!\!\exp \left[-\frac{1}{2b^{2}}\int_{0}^{N}
    \left(\frac{\partial{\mathbf{R}}}{\partial s}\right)^{2}\,\d s\right].
\end{equation}
The argument of the exponential is equivalent to the transversal 
kinetic energy, or more precisely to the 
effective action of the system since we integrate over the ``time'' 
$s$. Furthermore, this path integral obeys the classical
diffusion equation which arises  naturally from a pure RW process:
\begin{equation}
	\frac{\partial}{\partial s}Z(\mathbf{R},s)=
	\frac{b^{2}}{2}\frac{\partial^{2}}{\partial R^{2}}Z(\mathbf{R},s),
\end{equation}
with the boundary condition $Z(\mathbf{R},0)=\delta(\mathbf{R})$. 
For a forced RW, we introduce a potential term $\Phi(\mathbf{R}(s))$ 
standing for an external (virtual) strength field, in such a way that the 
total effective action takes the form 
\begin{equation}
	\beta\mathcal{\widehat{S}}=\frac{1}{2b^{2}}\int_{0}^{N}
	\left(\frac{\partial\mathbf{R}(s)}{\partial s}\right)^{2}\ {\rm{d}}s +
	\int_{0}^{N}\Phi(\mathbf{R}(s)){\rm{d}}s.
	\label{hamil}
\end{equation}
We note $\widehat{S}$ to emphasize the fact that this quantity represents a 
pseudo action (the integration is made over a Hamiltonian and not over a 
Lagrangian). This yields to the total partition function of the system, as 
the Feynman-Kac path integral:
\begin{equation}
	Z(\mathbf{R},s)=\mathcal{N}\int
	{\rm{exp}}\Big[-\beta\mathcal{\widehat{S}}\Big]\,\mathcal{D}\mathbf{R}(s).
\end{equation}
It is then convenient to introduce the transversal coordinate 
${\bf{y}}\equiv {\bf{R}}/b$ and so to write the action in the form 
\cite{dp2}
\begin{equation}
    \label{S1}
    \mathcal{\widehat{S}}=\int_{0}^{s}{\rm{d}}s'\left(\frac{1}{2}\Big(
    \frac{\d \mathbf{y}}{\d s'}\Big)^{2}+ 
    \Phi(\mathbf{y},s')\right)=
    \int_{0}^{s}{\rm{d}}s' H({\bf{y}},s').
\end{equation}
$H$ standing for the Hamiltonian of the system. By this way, we obtain the 
well-know expression of the partition function of directed polymers  
\begin{equation}
        \label{ZZ}
	Z(\mathbf{y},s)=\mathcal{N}\int_{\mathbf{y}(0)}^{\mathbf{y}(s)}
	\!\!\!\!\!\!\!\!\!\!\mathcal{D}\mathbf{y}\ {\rm{exp}}\left[-\beta
	\int_{0}^{s}{\rm{d}}s'\left(\frac{1}{2}\Big(
	\frac{\d \mathbf{y}}{\d s'}\Big)^{2}+ 
	\Phi(\mathbf{y},s')\right)\right];
\end{equation}
expression directly issued from a RW process with a directed line 
formalism. For a one dimensional system, $Z(y,s)$ obeys the following PDE 
\cite{kleinert}:
\begin{equation}
    \label{aa3}
    \frac{\partial}{\partial s}Z(y,s)=
    \nu\frac{\partial^{2}Z(y,s)}{\partial y^{2}}-\beta\Phi(y,s)\,Z(y,s),
\end{equation}
where $\nu=1/2\beta$ is a viscosity coefficient.

Let us underline that there exists several versions of the Hamiltonian 
depending on system to study \cite{ham1}-\cite{ham3}, 
\cite{ham4}-\cite{ham7}.

In view of these results, several analogies may be drawn between directed lines 
and vibrating strings. In its discrete form, the directed line is composed 
of a succession of rigid chains and the whole evolving with a resulting motion 
depending on the form of the noisy potential. For the continuous form, the term 
of inertia in polymer physics is due to the thermal agitation of the medium, and 
the  potential to an external field ({\it{e.g.}} electric field). In our case, 
the external potential is necessary because of the strength which have to be 
imposed to the line, in order to obtain realistic profiles of surface. So, 
the noisy potential of our model will be composed of a harmonic part and of a 
stochastic part which shall make the oscillations stochastic: 
$\Phi(y,s)=\frac{1}{2}ky^{2}+\eta(y,s)$, where $\kappa$ is the elastic 
constant of the line. The next step is to determine the equation of motion 
with the frequency of vibration of the line. In order to do this, 
let us proceed as follow:

In one hand, starting from Eq. (\ref{aa3}), it is well-known that we 
may obtain successively a KPZ and a Burgers equation (approach (A)):
\begin{equation}
    \label{res1}
    \frac{\partial\widehat{S}}{\partial s}=\nu\frac{\partial^{2}\widehat{S}}
    {\partial y^{2}}+\frac{1}{2}\left(\frac{\partial\widehat{S}}{\partial 
    y}\right)^2-\Phi;
\end{equation}
and
\begin{equation}
     \label{res2}
     \frac{\partial u}{\partial s}+u\frac{\partial u}{\partial y}=
     \nu\frac{\partial^{2} u}{\partial y^{2}}+\frac{\partial \Phi}{\partial y}.
\end{equation}
In this case the transformations are
\begin{equation}
    \label{res3}
    u=-\frac{\partial\widehat{S}}{\partial y}\quad{\rm{and}}\quad
    Z=\exp\left(\widehat{S}/2\nu\right).
\end{equation}
Usually one introduce the free energy. But the free energy tends to the total 
energy of the system $\widehat{S}$ when the entropy tends to zero.

In the other hand, starting from a general Burgers equation for a 
velocity field $u$, we have (approach (B))
\begin{equation}
     \label{res4}
     \frac{\partial u}{\partial s}+u\frac{\partial u}{\partial y}=
     \nu\frac{\partial^{2} u}{\partial y^{2}}-
     \frac{\partial V}{\partial y},
\end{equation}
and
\begin{equation}
    \label{res5}
    \frac{\partial F}{\partial s}=\nu\frac{\partial^{2}F}
    {\partial y^{2}}+\frac{1}{2}\left(\frac{\partial F}{\partial 
    y}\right)^2+V,
\end{equation}
and finally, with $\widehat{Z}=\exp\left(F/2\nu\right)$,
\begin{equation}
    \label{res6}
    \frac{\partial}{\partial s}\widehat{Z}(\mathbf{x},s)=
    \nu\nabla^{2}\widehat{Z}(\mathbf{x},s)+\frac{V}{2\nu}.
\end{equation}
Whether $u$ represents the transversal velocity, we have in this case
\begin{equation}
    \label{res7}
    u=\frac{\d y}{\d s}=-\frac{\partial F}{\partial y}.
\end{equation}
From these two approaches, a question arises: ``May the quantity $u$ of Eq. 
(\ref{res2}) be interpreted as the transversal velocity of Eq. (\ref{res4}) ?'' 
In other words, is there identity between $\widehat{S}$ and $F$ ?

To remove this ambiguity, let us start from the Burgers equation 
(\ref{res4}), $u$ given by (\ref{res7}), and let us use the relation
\begin{equation}
    \label{dem2}
    \d F(y,s)=\frac{\partial F}{\partial s}\d s+\frac{\partial 
    F}{\partial y}\d y;
\end{equation}
which yields to
\begin{equation}
    \label{dem3}
    \d F=\nu\frac{\partial^{2}F}{\partial y^{2}}\d s+
    \frac{1}{2}\left(\frac{\partial F}{\partial y}\right)^{2}\d s
    +V\;\d s+\frac{\partial F}{\partial y}\d y;
\end{equation}
and consequently to the expression
\begin{equation}
    \label{dem4}
    F=\nu\int\frac{\partial^{2}F}{\partial y^{2}}\d s+
    \frac{1}{2}\int\left(\frac{\partial F}{\partial y}\right)^{2}\d s
    +\int V\;\d s+\int\frac{\partial F}{\partial y}\d y.
\end{equation}
The first and last integrals may be rewritten 
\begin{equation}
    \label{dem5}
    \int\frac{\partial^{2}F}{\partial y^{2}}\d s=
    \frac{\partial}{\partial y}\int\frac{\partial F}{\partial y}\d s=
    -\frac{\partial}{\partial y}\int\frac{\d y}{\d s}\d s=-1.
\end{equation}
and
\begin{equation}
    \label{dem6}
    \int\frac{\partial F}{\partial y}\d y=
    -\int\frac{\d y}{\d s}\d y=-\int\left(\frac{\d y}{\d s}\right)^{2}\d s,
\end{equation}
In such a way that the ``free energy'' will read
\begin{equation}
    \label{dem7}
    F(y,s)=-\nu - \int_{0}^s 
    \Big[\frac{1}{2}\left(\frac{\d y}{\d s'}\right)^{2} 
    -V(y,s')\Big]\d s'\ .
\end{equation}
Finally, the Hopf-Cole transformation $\widehat{Z}=
\exp\left(F/2\nu\right)$, with $1/2\nu=\beta$, yields 
explicitly to
\begin{equation}
    \label{dem8}
    \widehat{Z}=e^{-1/2}\times{\rm{exp}}\left\{-\beta
    \int_{0}^{s}{\rm{d}}s'\left(\frac{1}{2}\left(
    \frac{\d y}{\d s'}\right)^{2}- V(y,s')\right)\right\}.
\end{equation}
The heat equation (\ref{aa3}) being invariant under the shift 
$Z\to\widehat{Z}=c\times Z$, $c\in\mathbb{R}$, we wiil have equivalence between 
the two approaches if $\Phi=-V$. And, therefore, one obtain the 
classical Hamilton equation of motion:
\begin{equation}
    \label{dem9}
    \frac{\d y}{\d s}=-\frac{\partial F}{\partial y}=
    -\frac{\partial\widehat{S}}{\partial y}\qquad
    \Longleftrightarrow\qquad\frac{\d^{2} y}{\d s^{2}}=
    -\frac{\partial H}{\partial y}
\end{equation}
The determination of an equation of motion for $y$ is henceforth equivalent to 
solve the KPZ equation (\ref{res1}). Moreover, we have a relation 
between the height of the growing surface and the transversal 
position: $|h(s)-h_{0}|\equiv |y(s)-y(0)|$, and so, by extension,
\begin{equation}
    \label{dem9}
    \frac{\d y}{\d s}\simeq\frac{\partial h}{\partial s}.
\end{equation}
For example $\partial h/\partial s=0\Leftrightarrow\d y/\d s=0$. Thereby, the 
action may as well be written \cite{h1,h2}
\begin{equation}
    \label{dem10}
   \mathcal{\widehat{S}}=
   \int_{0}^{s}\left(\frac{1}{2}\left(\frac{\partial h}{\partial s'}\right)^2+
    \Phi(h,s')\right)\d s',
\end{equation}

We shall now concentrate on the solution of the KPZ equation 
(\ref{res1}).

Analytic solutions of noisy KPZ equations have already been studied in the 
literature in specific cases \cite{exkpz1}-\cite{exkpz3}. But the 
non-linearity of this equation added to the presence of a stochastic noise 
make the resolution difficult. Nevertheless, it is known that this equation 
may be transformed either into a Burgers equation or a 
Schr$\rm{\ddot{o}}$dinger equation with an imaginary time. And, among works 
which treat exact solutions of theses kind of equations, we can cite 
\cite{st1}-\cite{moreau}. This is our concern here to treat the case of a 
line undergoing the stochastic elastic forcing term $-\kappa y+\eta(s)$, where 
the function $\eta$ contains all the noise that affects the oscillations 
({\it{e.g.}} a Langevin force).  This amounts to saying that we solve the KPZ 
equation (\ref{res1}) with a potential $\Phi$ given by 
$\Phi(y,s)=\frac{\kappa}{2}y^{2}-y\eta(s)$. Moreover, we assume a constant 
initial condition for the transversal velocity: $u(y,0)=u_{0}$.

\begin{figure}[!h]
	 \centerline{\includegraphics[scale=0.34]{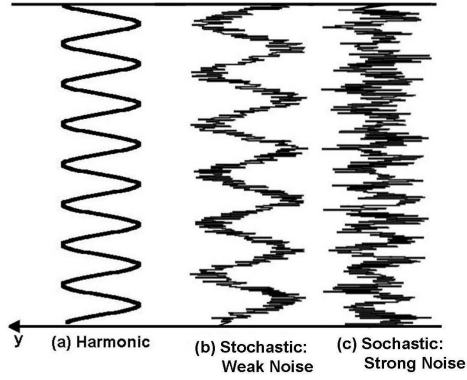}}
	 \caption{{\small{Variations of the transversal position $y$ for a 
	 forcing term inducing oscillations depending on the intensity of 
	 the noise. Each mode (a), (b), and (c) leads to very different 
	 roughness.}}}
	 \label{MHH}
\end{figure}

Following  methods presented in Ref. \cite{moreau}, it is shown in 
appendix that the solution is of the form:
\begin{equation}
    \label{sol1}
    \widehat{S}(y,s)=\frac{\xi(s)}{2}y^{2}-y\psi(s)+\Theta(s),
\end{equation}
where $\xi(s)$, $\psi(s)$ and $\Theta(s)$ are functions 
of $s$ and of $\eta(s)$ (see the appendix). But, we have 
the relation
\begin{equation}
    \label{sol2}
    \frac{\d y}{\d s}=-\frac{\partial\widehat{S}}{\partial y}=
    -y\xi(s)+\psi(s).
\end{equation}
Consequently, we obtain an equation standing for the equation of a forced 
oscillator with a term of friction:
\begin{equation}
    \label{sol3}
    \frac{\d^{2} y}{\d s^{2}}+\xi(s)\frac{\d y}{\d s}+\omega^{2}(s)\;y=F(s),
\end{equation}
where the pulsation $\omega^{2}$ and the forcing term $F$ are respectively 
given by $\d\xi/\d s$ and $\d\psi/\d s$. We can notice  from the expressions 
of $\xi$ and $\psi$ that, while the frequency $f\propto \sqrt{\d\xi/\d 
s}$ depends only on $s$, the term of force depends also on the stochastic noise 
$\eta$. Thus, we expect profiles of the form drawn in Fig. (\ref{MHH}).

Once $y$ determined, we can deduce the hvv from Eq. (\ref{sol2}):

\begin{equation}
    \label{sol4}
    \frac{\partial h}{\partial s}\simeq
    -\xi(s)\exp\left(-\int_{0}^s\xi(t)\d t\right)\left\{1+
    \int_{0}^s\psi(t)\exp\left(\int_{0}^t\xi(t')\d t'\right)\d 
    t\right\}
    +\psi(s).
\end{equation}
Then, the degree of irregularity of a surface, and so its global 
roughness, may be estimated by adding up all the local slopes 
$|\partial h/\partial s|$, in such a way that the interesting quantity to 
determine is $\langle |\partial h/\partial s|\rangle$, where the mean is taken 
over all the length of the surface. 

Note that this problem is different of the one of a stochastic oscillator, 
which is a specific problem \cite{stocho1,stocho2}
\section{Long size substrate}
We treat now the case of a particle of mass unity, moving along a 
rough surface when we study its profile at a fixed time. 

\begin{figure}[!h]
	 \centerline{\includegraphics[scale=0.34]{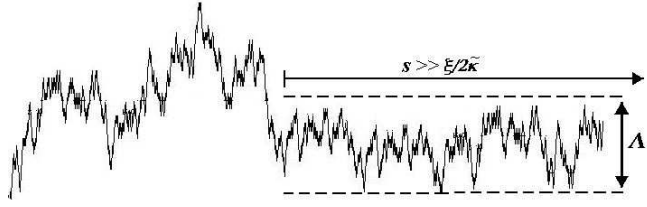}}
	 \caption{{\small{Asymptotic profile of surface for an O-U process. 
	 $\Lambda$ has to be interpreted as the maximal range of 
	 fluctuations.}}}
	 \label{MH3}
\end{figure}

Let us consider the propagation of a Brownian particle through virtual 
scatterers. Its trajectory describing the surface to analyse. We assume 
that the particle obeys a general Langevin equation  where an 
external force $\ -\partial\Phi/\partial y=-\tilde{\kappa} y$ is considered: 
\begin{equation}
    \label{beck1}
    \frac{\d^{2}y}{\d s^{2}}=-\xi\frac{\d y}{\d s}-
    \tilde{\kappa} y+\Gamma(s).
\end{equation}
This means that the particle undergoes effects of the medium (term 
of friction) in addition with an oscillator term. Then, the Langevin force 
$\Gamma(s)$ may be expressed in term of a Gaussian with noise $\eta(s)$, such 
as 
\begin{equation}
    \label{beck4}
     \frac{\d^{2}y}{\d s^{2}}=-\xi\frac{\d y}{\d s}-
    \tilde{\kappa}\;y+\sqrt{2\mathcal{D}}\eta(s);
\end{equation}
where $\mathcal{D}=\xi^{2}\tilde{\nu}$. By nature, the solution of this equation 
follows a Markovian process. When the inertial term is  negligible, 
one obtain  an Ornstein-Uhlenbeck process for the transversal position 
$y$:
\begin{equation}  
    \label{beck5}
    \frac{\d y}{\d s}=
    -\frac{\tilde{\kappa}}{\xi}y+\sqrt{2\tilde{\nu}}\; \eta(s).
\end{equation}
This well-known process \cite{orn1,orn2} is associated to the following 
Fokker-planck equation for the transition probability 
$\widetilde{P}(y,s|y',s')$:
\begin{equation}
    \label{mark5}
    \frac{\partial \widetilde{P}}{\partial s}=
    \frac{\tilde{\kappa}}{\xi}\frac{\partial y\widetilde{P}}{\partial y}+
    \tilde{\nu}\frac{\partial^{2}\widetilde{P}}{\partial y^{2}}.
\end{equation}
With the standard initial condition 
$\widetilde{P}(y,s|y_{0},s_{0})=\delta(y-y')$, 
the solution writes \cite{risken} 
\begin{equation}
    \label{mark6}
    P(y,y',s)=
    \sqrt{\frac{\tilde{\kappa}}{2\pi \xi\tilde{\nu}
    \left(1-e^{-\left(2\tilde{\kappa}/\xi\right)s}\right)}}\exp
    \left[-\frac{\tilde{\kappa}
    \big(y-e^{-\left(2\tilde{\kappa}/\xi\right)s}y'\big)^{2}}
    {2\xi\tilde{\nu}\big(1-e^{-\left(2\tilde{\kappa}/\xi\right)s\big)}}\right].
\end{equation}
This probability is asymptotically ``stationary'':
\begin{equation}
    \label{mark7}
    \lim_{s\to\infty}\widetilde{P}(y,y',s)=
    \sqrt{\frac{\tilde{\kappa}}{2\pi \xi\tilde{\nu}}}
    \exp\left[-\frac{\tilde{\kappa} y^{2}}{2\xi\tilde{\nu}}\right].
\end{equation}
In practice, the condition $s\gg\xi/2\tilde{\kappa}$ is sufficient to obtain 
this solution. This implies that, under this condition, the 
fluctuations of the transversal position $y$ becomes independent of the 
curvilinear abscissa. Moreover, the quantity 
$\Lambda=\sqrt{\xi\tilde{\nu}/\tilde{\kappa}}$, 
appearing in Eq. (\ref{mark7}) with the dimension of a length, takes the sense 
of the mean range for the transversal fluctuations with respect to a given 
position:
\begin{equation}
    \label{mark8}
    \int_{-\infty}^{+\infty}y^{2}\widetilde{P}(y)\d y=\Lambda^{2}.
\end{equation}
And the Einstein relation $\tilde{\nu}=1/\xi\beta$ gives us the expression
\begin{equation}
    \label{mark9}
    \Lambda\propto\tilde{\kappa}^{-1/2}
\end{equation}
Thereby, the higher the elastic constant, the lower the range of 
fluctuations.  The connection between the O-U process and the roughness is 
then highlighted in Fig. (\ref{MH3}), where we can notice that, for $s\gg 1$, 
the maximal fluctuation values of the height $h(y,s)$ are of the order of 
$\Lambda$. As a consequence, the hvv will be such as 
\begin{equation}
    \label{mark10}
    \left\arrowvert\frac{\partial h}{\partial s}\right\arrowvert\le
    \Lambda\qquad\forall s\gg \xi/2\tilde{\kappa}.
\end{equation}    
Consequently, in this limit,
\begin{equation}
    \label{c}
    \langle\left\arrowvert\frac{\partial h}{\partial s}
    \right\arrowvert\rangle\le\Lambda;
\end{equation}
available for models concerning processes of deposition executed from the left 
to the right of the figure, and on a ``large'' length of substrate.

Then, an interesting point lies in a connexion between the O-U process and  
KPZ equations. Indeed, the transformation 
\begin{equation}
    \label{mark11}
    \widetilde{P}(y,s)=\widetilde{Z}(y,s)e^{-\tilde{\kappa} y^{2}/4\tilde{\nu}},
\end{equation}
applied to Eq. (\ref{mark5}), leads to the heat equation
\begin{equation}
    \label{mark11}
    \frac{\partial\widetilde{Z}}{\partial s}=
    \tilde{\nu}\frac{\partial^{2}\widetilde{Z}}{\partial y^{2}}+
    \left(\frac{\tilde{\kappa}}{2}-
    \frac{\tilde{\kappa}^{2}y^{2}}{4\tilde{\nu}}\right)\widetilde{Z},
\end{equation}
and so to the KPZ equation
\begin{equation}
    \label{mark12}
    \frac{\partial\widetilde{S}}{\partial s}=
    \tilde{\nu}\frac{\partial^{2}\widetilde{S}}{\partial y^{2}}+
    \frac{1}{2}\left(\frac{\partial\widetilde{S}}{\partial y}\right)^{2}+    
    \left(\frac{\tilde{\kappa}}{2}-
    \frac{\tilde{\kappa}^{2}y^{2}}{4\tilde{\nu}}\right).
\end{equation}
And once more, the introduction of a quantity $u$ such as 
$u=-\partial\widehat{S}/\partial y$, yields to a Burgers equation:
\begin{equation}
    \label{mark13}
    \frac{\partial u}{\partial s}+u\frac{\partial u}{\partial y}=
    \tilde{\nu}\frac{\partial^{2}u}{\partial y^{2}}-
    \frac{\partial\tilde{\Phi}}{\partial y}.
\end{equation}
But the same interrogation as the one of the RW process may be done, 
namely, may $u$ be considered as the velocity $\d y/\d s$ of Eq. 
(\ref{beck5}) ?  
As a matter of fact, these relations have been obtained with the 
assumption $\d^{2}y/\d s^{2}\simeq 0$. So, let us assume that $u=\d y/\d s$. 
The consequence is that the first term of Eq. (\ref{mark13}) vanishes, and so 
we obtain
$$\frac{1}{2}\frac{\partial\tilde{u}^{2}}{\partial y}=
\tilde{\nu}\frac{\partial^{2}\tilde{u}}{\partial y^{2}}-
\frac{\partial\tilde{\Phi}}{\partial y}$$ 
\begin{equation}
    \label{mark14}
    \Leftrightarrow \frac{\partial}{\partial y}\left\{
    \frac{1}{2}\left(\frac{\d y}{\d s}\right)^{2}+\tilde{\Phi}\right\}=
    \frac{\partial H}{\partial y}=\tilde{\nu}\frac{\partial^{2}}{\partial y^{2}}
    \left(\frac{\d y}{\d s}\right).
\end{equation}    
Furthermore, we have that $-\partial H/\partial y
=\d^{2}y/\d s^{2}\simeq 0$. So the last part of Eq. (\ref{mark14}), leads 
to a solution of the form     
\begin{equation}
    \label{mark15}
    \frac{\d y}{\d s}=c_{1}(s)y+c_{2}(s).
\end{equation}
Thus, $u=\d y/\d s\Leftrightarrow$ $u$ obeys to Eq. (\ref{mark15}). Then, 
we refound Eq. (\ref{beck5}) by putting $c_{1}=-\tilde{\kappa}/\xi$ and 
$c_{2}(s)=\sqrt{2\tilde{\nu}}\eta(s)$.

Once again, a stochastic process (here Markovian) may be connected to 
the coupled KPZ/Burgers equations, and by extension to the Hamiltonian of the 
system. In addition, a partition function like the one of the RW may be 
obtained with appropriated $\tilde{\nu}$ and $\tilde{\kappa}$. 
Moreover, this shown the relevant choice of a harmonic potential $\sim y^{2}$ 
to treat these kind of fluctuations, since a Markovian 
process and a RW process yields both to the same class of equations.
\section{Conclusion}
Two short stochastic models of directed line have been presented and applied 
to the description of rough surfaces. The background of our approach was that 
roughness of a one-dimensional surface is created by the motion of a virtual 
particle dived in  a stochastic harmonic potential. It brings out of the 
continuous RW that we can rigorously linked the partition function of the 
line to the Hamilton equation of motion for the transversal position. The 
survey of roughness amounts therefore to solve a KPZ equation with a particular 
potential. Then, interpreting the roughness as a result of a 
stochastic elastic forcing term acting on a surface, we have obtained a 
forced oscillator equation where the force behaves as a noise for the line. 
The analytical expressions of the pulsation and of the force 
have been determined too. Consequently, local analysis of the roughness may be 
performed thanks to the expression of the hvv (\ref{sol4}) which, besides, 
obeys a Burgers equation. For a sufficient long substrate, the O-U model shows 
that the maximal values of roughness are asymptotically bounded in a range of 
length $\Lambda\propto\tilde{\kappa}^{-1/2}$. The elastic constant being an 
adjustable parameter mastering the magnitude of the fluctuations. 
Moreover, we have shown that the two models, used with a same kind of force, 
had common properties since they lead both to the same class of 
equations. As a matter of fact, the two equations of motion are connected to 
the partition function of the system. Finally, these two models belong to 
stochastic processes described by the Burgers dynamics.

Although we have presented one-dimensional models, these approaches could 
open interesting ways of investigation for numerical and analytical 
studies, in particular if one needs to generate surfaces with specific 
profiles or properties.
\newpage
\appendix*
\section{Solution of the KPZ equation}

We solve here the KPZ equation (\ref{res1}) or, equivalently, the heat 
equation (\ref{aa3}) written in the following way:
\begin{equation}
    \label{ap0}
    \frac{\partial}{\partial s}Z=\nu\frac{\partial^{2}Z}{\partial y^{2}}+
    \left(-\frac{\beta\kappa}{2}y^{2}+\beta\eta\;y\right)Z,
\end{equation}
with the initial condition $Z(y,0)=\exp\left(-u_{0}y/2\nu\right)$. 
The resolution is based on the Time Space Transformation method (TST) 
of \cite{moreau}. We give only the main points of resolution. We 
would like to underline for the comprehension that the notations $x$ and 
$y$ are reversed between the present paper and \cite{moreau}: 
$y\leftrightarrow x$.

The first step is to determine the initial condition. For this, we need 
to determine the expression for the variables $a_{1}$, $a_{2}(s)$, $p(s)$, 
$q(s)$ and $\tau(s)$ of \cite{moreau}. We obtain successively:
\begin{equation}
    \label{ap1}
    a_{1}=\sqrt{\kappa}/4\nu,
\end{equation}
\begin{equation}
    \label{ap2}
    a_{2}(s)=e^{-\sqrt{\kappa}s}\left(1+
    \beta\int_{0}^s \eta(t)e^{\sqrt{\kappa}t}\d t\right),
\end{equation}
\begin{equation}
    \label{ap3}
    p(s)=e^{\sqrt{\kappa}s},
\end{equation}
\begin{equation}
    \label{ap4}
    q(s)=2\nu s+\int_{0}^s \int_{0}^t \eta(t') 
    e^{\sqrt{\kappa}t'}\d t' \d t,
\end{equation}
\begin{equation}
    \label{ap5}
    \tau=e^{\sqrt{\kappa}s}\sinh(\sqrt{\kappa}s)/\sqrt{\kappa}.
\end{equation}
In such a way that the resolution of (\ref{ap0}) amounts to solve the 
equation:
\begin{equation}
    \label{ap6}
    \frac{\partial\Phi}{\partial\tau}=
    \nu\frac{\partial^{2}\Phi}{\partial x^{2}},
\end{equation}
with the initial condition 
\begin{equation}
    \label{ap7}
    \Phi(x,0)=\exp\left(-a_{1}x^{2}-(1+u_{0}/2\nu)x\right). 
\end{equation}
The solution reads explicitly
\begin{equation}
    \label{}
    \Phi(x,\tau)=
    (4a_{1}\nu\tau+1)^{-1/2}\exp\left\{-a_{1}x^{2}-(1+u_{0}/2\nu)x+
    \frac{\nu\tau(2a_{1}x+1+u_{0}/2\nu)^{2}}{(`a_{1}\nu\tau+1)}\right\}.
\end{equation}
Then, reversing the different transformations, we obtain as a solution of 
(\ref{ap0})
$$  Z(y,s)=\exp\left\{y^{2}\left(a_{1}-a_{1}e^{2\sqrt{\kappa}s}+
    \frac{4a_{1}^{2}\nu}{\sqrt{\kappa}}e^{2\sqrt{\kappa}s}
    \tanh(\sqrt{\kappa}s)\right)\right.+
$$

$$
    y\left(a_{2}(s)-2a_{1}q(s)e^{\sqrt{\kappa}s}-
    (1+u_{0}/2\nu)e^{\sqrt{\kappa}s}+
    \frac{2\nu a_{1}}{\sqrt{\kappa}}(1+u_{0}/2\nu+
    2a_{1}q(s))e^{\sqrt{\kappa}s}\tanh(\sqrt{\kappa}s)\right)+
$$
\begin{equation}
    \left.\left(-a_{1}q^{2}(s)+\frac{\nu}{\sqrt{\kappa}}(1+2a_{1}q(s))^{2}
    \tanh(\sqrt{\kappa}s)-\frac{\sqrt{\kappa}}{2}s-\frac{1}{2}
    \ln\left(\cosh(\sqrt{\kappa}s)\right)\right)\right\}
\end{equation}
expression that we write for further analyse
\begin{equation}
    \label{}
    Z(y,s)=\exp\left(\frac{\xi(s)}{4\nu}y^{2}-\frac{\psi(s)}{2\nu}y+
    \frac{\Theta(s)}{2\nu}\right),
\end{equation}
where
\begin{equation}
    \label{}
    \xi(s)=4\nu\left(a_{1}-a_{1}e^{2\sqrt{\kappa}s}+
    \frac{4a_{1}^{2}\nu}{\sqrt{\kappa}}e^{2\sqrt{\kappa}s}
    \tanh(\sqrt{\kappa}s)\right),
\end{equation}
\begin{displaymath}
    \label{}
    \psi(s)=-2\nu
    \left(a_{2}(s)-2a_{1}q(s)e^{\sqrt{\kappa}s}-
    (1+u_{0}/2\nu)e^{\sqrt{\kappa}s}+
    \frac{2\nu a_{1}}{\sqrt{\kappa}}(1+u_{0}/2\nu+\right.
\end{displaymath}
\begin{equation}
    \left.+2a_{1}q(s))e^{\sqrt{\kappa}s}\tanh(\sqrt{\kappa}s)\right),
\end{equation}
\begin{equation}
    \label{}
    \Theta(s)=2\nu\left(-a_{1}q^{2}(s)+
    \frac{\nu}{\sqrt{\kappa}}(1+2a_{1}q(s))^{2}
    \tanh(\sqrt{\kappa}s)-\frac{\sqrt{\kappa}}{2}s-\frac{1}{2}
    \ln\left(\cosh(\sqrt{\kappa}s)\right)\right),
\end{equation}
The quantity $a_{1}$, $a_{2}(s)$ and $q(s)$ being respectively given by 
Eqs. (\ref{ap1}), (\ref{ap2}) and (\ref{ap4}). Note that we recover the 
initial condition $Z(y,0)$. Then, since $Z=\exp\left(\widehat{S}/2\nu\right)$, 
we can write the solution on the final form 
\begin{equation}
    \label{}
    \widehat{S}(y,s)=\frac{\xi(s)}{2}y^{2}-y\psi(s)+\Theta(s)
\end{equation}


\begin{thebibliography}{120}

\bibitem{ex1}
\textsc{P. Brault, A-L. Thomann, C. Andreazza-Vignolle}, Surf. Sci.\textbf{406}, 
L597-L602 (1998). And references therein.

\bibitem{ex2}
\textsc{S.M. Rossnagel, J.J. Cuomo, W.D. Westwood}, Handbook of Plasma Processing 
Technology, Noyes Publications, Park Ridge, NJ, 1990.

\bibitem{ex3}
\textsc{J.L. Viovy, D. Beysens and C.M. Knobler}, Phys. Rev. A 37 (1988) 4965.

\bibitem{ex4}
\textsc{G.S. Bales and D.C. Chrzan}, Phys. Rev. B 50 (1994) 6057.

\bibitem{kpz1}
\textsc{M. Kardar, G. Parisi, and Yi-Cheng Zhang}, Phys. Rev. Lett. 
\textbf{56}, 889 (1986). 

\bibitem{kpz2}
\textsc{F. Ginelli and H. Hinrichsen},J. Phys. A \textbf{37} 11085 (2004).

\bibitem{kpz3}
\textsc{Y. Kim and S. H. Yook}, J. Phys. A \textbf{30} L449 (1997) .

\bibitem{kpz4}
\textsc{T. J. Newman and A. J. Bray}, J. Phys. A \textbf{29} 7917 (1996).

\bibitem{kpz5}
\textsc{Z. Csah$\grave{\rm{o}}$k, K. Honda, and T. Vicsek}, J. Phys. A  
\textbf{26} L171 (1993). 

\bibitem{kpz6}
\textsc{R.G. da Silva, M.L. Lyra, C.R. da Silva, and G.M. Viswanathan}, 
Eur. Phys. J. B \textbf{17}, 693 (2000).

\bibitem{kpz7}
\textsc{T. J. Newman and H. Kallabis}, J. Phys. I France \textbf{6} 373  (1996).

\bibitem{kpz8}
\textsc{Deok-Sun Lee and D. Kim}, J. Stat. Mech. P08014 (2006).

\bibitem{kpz9}
\textsc{S. V. Ghaisas}, Eur. Phys. J. B \textbf{52}, 557 (2006).

\bibitem{kpz10}
\textsc{M. Myllys, J. Maunuksela, J. Merikoski, J. Timonen, and M. 
Avikainen}, Eur. Phys. J. B \textbf{36}, 619 (2003).

\bibitem{kpz11}
\textsc{Z. R$\acute{\rm{a}}$cz}, arXiv:cond-mat/0307490

\bibitem{ham1}
\textsc{E. B. Kolomeisky and J. P. Straley}, Phys. Rev. B \textbf{46}, 12664 
(1992).

\bibitem{ham2}
\textsc{R. Lipowsky and H. Müller-Krumbhaar}, Europhys. Lett. \textbf{11}, 657 
(1990).

\bibitem{ham3}
\textsc{G. Parisi and F. Slanina}, Eur. Phys. J. B \textbf{8}, 603 (1999).

\bibitem{attention 2 fois}
\textsc{M. Kardar and Y-C. Zhang}, Phys. Rev. Lett. \textbf{58}, 2087 (1987).

\bibitem{ham4}
\textsc{D. A. Huse, W. Van Saarloos, and J. D. Weeks}, Phys. Rev. B 
\textbf{32}, 233 (1985).

\bibitem{ham5}
\textsc{D. A. Huse and C. L. Henley}, Phys. Rev. Lett.\textbf{54}, 2708 (1985).

\bibitem{attention 2fois}
\textsc{H.K. Janssen1, U.C. Tauber, and E. Frey}, Eur. Phys. J. B 9, 491511 
(1999).

\bibitem{ham6}
\textsc{G. Blatter, M. V. Feigel'man, V. G. Geschkenbein, A. I. Larkin and 
V. M. Vinokur}, Rev. Mod. Phys. \textbf{66}, 1125 (1994).

\bibitem{ham7}
\textsc{M. Kardar}, Phys. Rev. Lett. \textbf{55}, 2235 (1985).

\bibitem{dp1}
\textsc{T. A. Vilgis}, Phys. Rep. \textbf{336}, 167 (2000).

\bibitem{dp2}
\textsc{T. J. Newman and A. J. McKane}, Phys. Rev. E \textbf{55}, 165 (1997).

\bibitem{dp3}
\textsc{Y. Kifer}, Probab. Theory Relat. Fields \textbf{108}, 29 (1997).

\bibitem{dp4}
\textsc{A. Bulgac and H. E. Roman}, New Journal of Physics \textbf{7}, 2 (2005).

\bibitem{fp1}
\textsc{A. Bovier and J. Fr$\ddot{o}$hlich}, Phys. Rev. B \textbf{34}, 6409 
(1986).

\bibitem{fp2}
\textsc{M. Kardar and Y-C. Zhang}, Phys. Rev. Lett. \textbf{58}, 2087 (1987).

\bibitem{fp3}
\textsc{D. A. Huse, C. L. Henley, and D. S. Fisher}, Phys. Rev. Lett. 
\textbf{55}, 2924 (1985). 

\bibitem{r1}
\textsc{S. B. Singha}, J. Stat. Mech. P08006 (2005).

\bibitem{r2}
\textsc{C.M. Horowitz and E.V. Albano}, Eur. Phys. J. B \textbf{31}, 563 (2003).

\bibitem{r3}
\textsc{J. Asikainen, S. Majaniemi, M. Dub$\acute{\rm{e}}$, J. Heinonen, and 
T. Ala-Nissila}, Eur. Phys. J. B \textbf{30}, 253 (2002).

\bibitem{r4}
\textsc{A. Hader, A. Memsouk, and Y. Boughaleb}, Eur. Phys. J. B \textbf{28}, 
315 (2002). 

\bibitem{quenched-noise}
\textsc{A. D$\acute{\rm{i}}$az-S$\acute{\rm{a}}$nchez, L.A. Braunstein and R.C. 
Buceta}, Eur. Phys. J. B \textbf{21}, 289 (2001).

\bibitem{r5}
\textsc{M. Dub$\acute{\rm{e}}$, M. Rost, and M. Alava}, Eur. Phys. J. B 
\textbf{15}, 691 (2000).

\bibitem{markov1}
\textsc{M. Waechter, F. Riess, Th. Schimmel, U. Wendt, and J. Peinke},
Eur. Phys. J. B \textbf{41}, 259 (2004).
\bibitem{markov2}
\textsc{G. R. Jafari, S.M. Fazeli, F. Ghasemi, S.M. Vaez Allaei, M. Reza Rahimi 
Tabar, A. Iraji zad, and G. Kavei}, Phys. Rev. Lett. \textbf{91}, 226101 (2003).

\bibitem{frac}
\textsc{L. Niemeyer, L. Pietronero, and H. J. Wiesmann}. \emph{Scaling 
properties of growing zone and capacity of Laplacian fractal}. 
Amsterdam: Elsevier (1986).


\bibitem{kleinert}
\textsc{H. Kleinert}, \emph{Path integrals in Quantum Mechanics, 
Statistics and Polymers Physics}, World Scientific (1995).

\bibitem{h1}
\textsc{S. Das Sarma, C. J. Lanczycki, R. Kotlyar and S. V. Ghaisas}, 
Phys. Rev. E \textbf{53}, 359 (1996).

\bibitem{h2}
\textsc{Z.-W. Lai and S. Das Sarmas}, Phys. Rev. Lett. \textbf{66}, 2348 (1991).

\bibitem{exkpz1}
\textsc{H.K. Janssen1, U.C. Tauber, and E. Frey}, Eur. Phys. J. B 9, 491 (1999).

\bibitem{exkpz2}
\textsc{T. J. Newman}, Phys. Rev. E \textbf{51}, 4212 (1995).

\bibitem{exkpz3}
\textsc{S. E. Esipov}, Phys. Rev. E, \textbf{49}, 2070 (1994).

\bibitem{st1}
\textsc{L. Bertini, N. Cancrini, and G. Jona-Lasinio}, Commun. Math. Phys. 
\textbf{165}, 211 (1994).

\bibitem{st2}
\textsc{O. Al Hammal, F. de los Santos, and M. A. 
Mu$\tilde{\rm{n}}$oz}, J. Stat. Mech. P10013 (2005).

\bibitem{st3}
\textsc{T. Laurila, C. Tong, I. Huopaniemi, S. Majaniemi and T. 
Ala-Nissila},  Eur. Phys. J. B \textbf{46}, 553 (2005).

\bibitem{st4}
\textsc{I. A. Davies, A. Truman, and H. Zhao}, J. Math. Phys. \textbf{46}, 
043515 (2005).

\bibitem{moreau}
\textsc{E. Moreau and O. Vallee}, Phys. Rev. E  \textbf{73}, 016112 (2006).  

\bibitem{stocho1}
\textsc{N. Van Kampen}, Stochastic processes in physics and chemistry, North 
Holland Personal Library, (2001).

\bibitem{stocho2}
\textsc{F. J. Poulin and M. Scott}, Nonlin. Processes Geophys. 
\textbf{12}, 871 (2005).

\bibitem{orn1}
\textsc{G. E. Uhlenbeck {\it{\&}} L. S. Ornstein} Phys. Rev. 
{\bf{36}}, 823 (1930).

\bibitem{orn2}
\textsc{M. C. Wang {\it{\&}} G. E. Uhlenbeck} Rev. Mod. Phys. 
{\bf{17}}, 323 (1945).


\bibitem{risken} 
\textsc{H. Risken}, \emph{The Fokker-Planck equation},
second edition, Spinger (1989).







\end{thebibliography}
\end{document}